\documentclass[12pt,preprint]{aastex}
\usepackage{natbib}
\usepackage{lscape}
\begin{document}
\slugcomment{ApJ, in press}
\title{HIP 38939B: A New Benchmark T Dwarf in the Galactic Plane Discovered with Pan-STARRS1}
\shorttitle{HIP 38939B}

\author{
Niall R. Deacon,\altaffilmark{1,2}~\footnote{deacon@mpia.de}
Michael C. Liu,\altaffilmark{1,3}
Eugene A. Magnier,\altaffilmark{1}
Brendan P. Bowler,\altaffilmark{1}
Joshua Redstone,\altaffilmark{4}
Bertrand Goldman,\altaffilmark{2}
W. S. Burgett,\altaffilmark{1}
K. C. Chambers,\altaffilmark{1}
H. Flewelling,\altaffilmark{1}
N. Kaiser,\altaffilmark{1}
J.S. Morgan,\altaffilmark{1}
P.A. Price,\altaffilmark{5}
W.E. Sweeney,\altaffilmark{1}
J.L. Tonry,\altaffilmark{1}
R.J. Wainscoat,\altaffilmark{1}
C. Waters,\altaffilmark{1}}
\altaffiltext{1}{Institute for Astronomy, University of Hawai`i, 2680 Woodlawn Drive, Honolulu, HI 96822, USA}
\altaffiltext{2}{Max Planck Institute for Astronomy, Koenigstuhl 17, D-69117 Heidelberg, Germany}
\altaffiltext{3}{Visiting Astronomer at the Infrared Telescope Facility, which is operated by the University of Hawaii under Cooperative Agreement no. NNX-08AE38A with the National Aeronautics and Space Administration, Science Mission Directorate, Planetary Astronomy Program}
\altaffiltext{4}{Facebook, 1601 S. California Avenue, Palo Alto, CA 94304e, USA}
\altaffiltext{5}{Princeton University Observatory, 4 Ivy Lane, Peyton Hall, Princeton University, Princeton, NJ 08544, USA}
 \label{firstpage}
 \begin{abstract}
 We report the discovery of a wide brown dwarf companion to the mildly metal-poor ([Fe/H]=-0.24), low galactic latitude ($b$=1.88$^\circ$) K4V star HIP 38939. The companion was discovered by its common proper motion with the primary and its red optical (Pan-STARRS1) and blue infrared (2MASS) colors. It has a projected separation of 1630 AU and a near-infrared spectral type of T4.5. As such it is one of only three known companions to a main sequence star which have early/mid-T spectral types (the others being HN Peg~B and $\epsilon$~Indi~B). Using chromospheric activity we estimate an age for the primary of 900$\pm^{1900}_{600}$ Myr. This value is also in agreement with the age derived from the star's weak ROSAT detection. Comparison with evolutionary models for this age range indicates that HIP 38939B falls in the mass range 38$\pm$20 M$_{\rm Jup}$ with an effective temperature range of 1090$\pm$60 K. Fitting our spectrum with atmospheric models gives a best fitting temperature of 1100 K.  We include our object in an analysis of the population of benchmark T dwarfs and find that while older atmospheric models appeared to over-predict the temperature of the coolest objects compared to evolutionary models, more recent atmospheric models provide better agreement. 
 \end{abstract}
 \section{Introduction}
The identification of a large sample of field L~and T~dwarfs over the
last $\approx$15~years has progressed from individual discoveries to
large samples of hundreds of objects. Yet in contrast to the tremendous
advances made in spectral classification \citep[e.g.][]{Kirkpatrick2005},
accurate determination of the physical properties of ultracool dwarfs
remains an unfulfilled goal. One inherent difficulty originates from the
fundamental nature of substellar objects. In the absence of sustained
internal energy generation, the temperatures, luminosities, and radii
for substellar objects of a given mass steadily decline with age ---
thus diagnosing physical properties from photometry and spectroscopy is
more complicated than for main-sequence stars, which maintain much more
constant properties over their lifetimes. In addition, theoretical
modeling of emergent fluxes and spectra at such low effective
temperatures ($\lesssim$2500~K) is challenged by uncertainties in the
treatment of dust, clouds, molecules, and chemistry.

To help overcome these difficulties, the rare subset of brown dwarfs
known as ``benchmarks'' is highly valued. This designation is commonly
used for any brown dwarf with some physical properties much better
constrained compared to field objects. Given that three fundamental
quantities (e.g., luminosity, temperature, and age) are needed to
primarily determine a brown dwarf's physical state (ignoring metallicity),  a sharper
definition would be any object for which two fundamental quantities are
well-constrained. As described in \citet{Liu2008}, substellar benchmarks
can be naturally distinguished into ``mass benchmarks'' and ``age
benchmarks'', where the mass or age, respectively, are well-constrained.
Mass benchmarks are ultracool binaries with directly determined
distances (and hence luminosities) and dynamical masses. In these binaries, the remaining main
properties (temperature and surface gravity) can be very precisely
constrained using evolutionary models. However, given the long orbital
timescales ($\gtrsim$10~yr), the current sample of mass benchmarks for
the L~and T~spectral classes is limited to five systems (e.g. see compilation
in \citealp{Dupuy2011}).\footnote{There is only one known
  field system that is both an age and mass benchmark, namely the L4+L4
  binary HD~130948BC, which has a directly dynamical mass determination
  and a well-constrained age for the primary star from gyrochronology
  \citep{Dupuy2008}.} Ample age benchmarks at hotter
temperatures (spectral types of late-M to early-L) have been found in
young ($\lesssim$1--100~Myr) clusters
\citep[e.g.][]{Moraux2003, Slesnick2006,Lodieu2008}, where the distances and stellar ages are very
well-known, and thus the luminosites and ages of the brown dwarfs are as
well. However, open clusters with ages comparable to the field
population ($\gtrsim$Gyr) are largely dynamically depleted of substellar
objects \citep[e.g.][]{Bouvier2008}.

Thus, the search for field benchmarks has largely focused on brown dwarfs
found as companions to stars (e.g., see compilation in
\citealp{Faherty2010}). In this case, the ages of the primary
stars can also be attributed to their brown dwarf companions, though
their typical age uncertainties lead to poorer constraints compared to
brown dwarfs in short-period binaries or open clusters. Three T-dwarf
companions have been found at small angular separations through adaptive
optics imaging, including the archetype Gl~229B \citep{Nakajima1995},
GJ~758B \citep{Thalmann2010}, and SCR~1845$-$6357B \citep{Biller2006}.
Nine other T~dwarfs have been found at much wider separations around
main-sequence stars from wide-field survey data: GL~570D
\citep{Burgasser2000}, HD~3651B \citep{Mugrauer2006},
HN~PegB\citep{Luhman2007}, Wolf~940B \citep{Burningham2009}, Ross~458C
\citep{Goldman2010}, HIP~63510C \citep{Scholz2010B}, HIP~73786B
(\citealt{Scholz2010B}, \citealt{Murray2011}), $\epsilon$~Indi~Bab
\citep{Scholz2003, McCaughrean2004}, Gl~337CD \citep{Wilson2001,Burgasser2005}. \cite{Albert2011} identify CFBDS J022644$-$062522 as a possible wide companion to HD15220 but while the secondary's spectroscopic parallax places it at a similar distance to the primary, the proper motion determination on the secondary is currently not accurate enough to confirm companionship. Additionally, LSPM 1459+0857B
\citep{Day-Jones2010} and WD~0806$-$661b \citep{Luhman2011} have been
identified as wide companions to white dwarfs, the latter of which is
much cooler than any known T~dwarfs but has not yet been spectrally
classified. 

Wide-field surveys represent fertile ground for identifying benchmarks.
Most wide benchmark companions have resulted from large-area searches
for free-floating ultracool dwarfs, which are then realized to be
comoving with higher mass stars. More limited efforts have been carried
out to expressly identify wide benchmarks around well-defined samples of primary stars
\citep{Pinfield2006,Day-Jones2008}. With a sufficiently
large sample of benchmarks, it may be possible to provide a direct
empirical calibration that ties the underlying physical parameters (e.g.,
age and temperature) to observable brown dwarf properties
(e.g. \citealt{Pinfield2006}). To this end, the Pan-STARRS1 (PS1)
Telescope \citep{Kaiser2010} is a welcome addition to help boost the
census of benchmark brown dwarfs.

Situated on the summit of Haleakal\={a} on the island of Maui in the
Hawaiian Islands, PS1 will be the leading multi-epoch optical survey
facility over the next several years. PS1 is the first of four planned
1.8-meter telescopes that will comprise the Panoramic Survey Telescope
And Rapid Response System (\citealt{Kaiser2002}). Full science
operations with PS1 began in May 2010, and fabrication of the second telescope (PS2) has
already begun. So far, commissioning and regular
survey data have been used to search for Trans-Neptunian Objects
(\citealt{Wang2009}), T~dwarfs (\citealt{Deacon2011}, \citealt{Liu2011})
and supernovae (e.g. \citealt{Botticella2010}). The telescope is
executing several astronomical surveys of varying depth, area, cadence,
and filter complements. The most interesting of these for discovering
free-floating and companion brown dwarfs is the 3$\pi$ Survey, which is
scanning the entire sky north of $\delta=-30^{\circ}$ (3$\pi$
steradians) in five filters ($g_{P1}$, $r_{P1}$, $i_{P1}$, $z_{P1}$ and
$y_{P1}$; \citealt{Stubbs2010}, \citealt{Tonry2011} in prep.) at six separate epochs over 3.5~years,
with each epoch consisting of pairs of exposures taken $\approx$25~min
apart. These multiple epochs and the use of the
0.95-1.03~\micron~$y_{P1}$ band make the survey well-suited for
identifying nearby cool objects such as brown dwarfs by their proper
motions and parallaxes (\citealt{Magnier2008}, \citealt{Beaumont2010}).

In this paper, we present the discovery of a T4.5~dwarf around the
K4V~star HIP~38939 found from a dedicated search for wide late-L and T dwarf companions
using the PS1 and 2MASS. In \S~2, we describe our mining of these
datasets to search for companions to main-sequence stars with proper
motions from Hipparcos. Spectroscopic confirmation is presented in \S~3.
In \S~4 we examine the physical properties of our new discovery, along
with a look at the total compilation of benchmark T~dwarfs now known.
 
\section{Identification in PanSTARRS1 + 2MASS data}
Pan-STARRS1 began full survey operations in May 2010. Hence for a given filter and location on the sky, the current 3$\pi$ survey data consist of only one pair of images taken at the same epoch or multiple pairs separated by a short time baseline. So we can use these data for studies of nearby brown dwarfs, we have undertaken a proper motion survey by combining PS1 data with those from 2MASS~\citep{Skrutskie2006}. This method was used to identify new T dwarfs in \cite{Deacon2011}.  We constructed a database of high proper motion objects containing all PS1 3$\pi$ survey commissioning and survey data taken from the beginning of February 2010 to the end of January 2011. This was done by identifying PS1 $y_{P1}$ objects with no 2MASS detection within one arcsecond. We required that the PS1 object was detected on at least two images, was classified as a good quality point source detection, and had a photometric error less than 0.2 magnitudes in  the $y_{P1}$ band. The double detection requirement excludes contamination due to asteroids and reduces the number of instrumental artefacts. These objects are then paired with 2MASS $J$-band point sources which had an A, B or C photometric quality flag designation (corresponding to S/N$>$5), which were within 28$\arcsec$ and which had no corresponding PS1 detection. This pairing radius was chosen to allow the detection of objects with proper motions up to 2$\arcsec$/yr given the 14-year maximum epoch difference between PS1 and 2MASS. 

To identify companion objects to nearby stars, we ran queries on our proper motion database around all stars in the Hipparcos catalogue \citep{Perryman1997}. We chose stars with proper motions above 0.1$\arcsec$/yr and parallaxes more significant than $5 \sigma$. We searched for companions with projected separations less than 10,000 AU at the distance of the Hipparcos star. Each candidate was further filtered to have proper motion vectors within $5 \sigma$ of the primary assuming a nominal accuracy of 16 milliarcseconds per year\footnote{This is a conservative estimate based on a comparison between the measured proer motions for stars in the \cite{Lepine2005} catalog and the proper motions we derived for these objects.} in each coordinate for our proper motions and errors from \cite{Perryman1997} for the primary. We then picked out candidate ultracool dwarfs using the color selections found in \cite{Deacon2011} which are sensitive to late-L and T dwarfs. These are $y_{P1} -J > 2.2$, $J - H < 1.0$ and $z_{P1} - y_{P1} > 0.6$ with no corresponding $R$ or $I$ band detections in SuperCOSMOS \citep{Hambly2001} or USNO-B \citep{Monet2003} brighter than 20.5 and 18.5 magnitudes respectively. No Galactic latitude cut was applied. Taking into account area lost due to gaps between chips in the PS1 camera and areas of multiple coverage where these gaps can be filled in, we searched 16,900 sq.deg. After examining 1008 candidates by eye to remove spurious associations and objects with clear but uncatalogued companions in the USNO-B photographic plate images 380 candidates remained. Of these, our clearest late-type candidate was a companion to HIP~38939 (a.k.a. CD-25~5342, GJ~9246, HD~65486, NLTT~18729, SAO~174889). We have not yet completely followed up our remaining candidates, hence this work to date should not be considered a complete survey for late-type companions to Hipparcos stars.

\begin{figure}[htbp]
\begin{center}
\epsscale{1.0}
\plotone{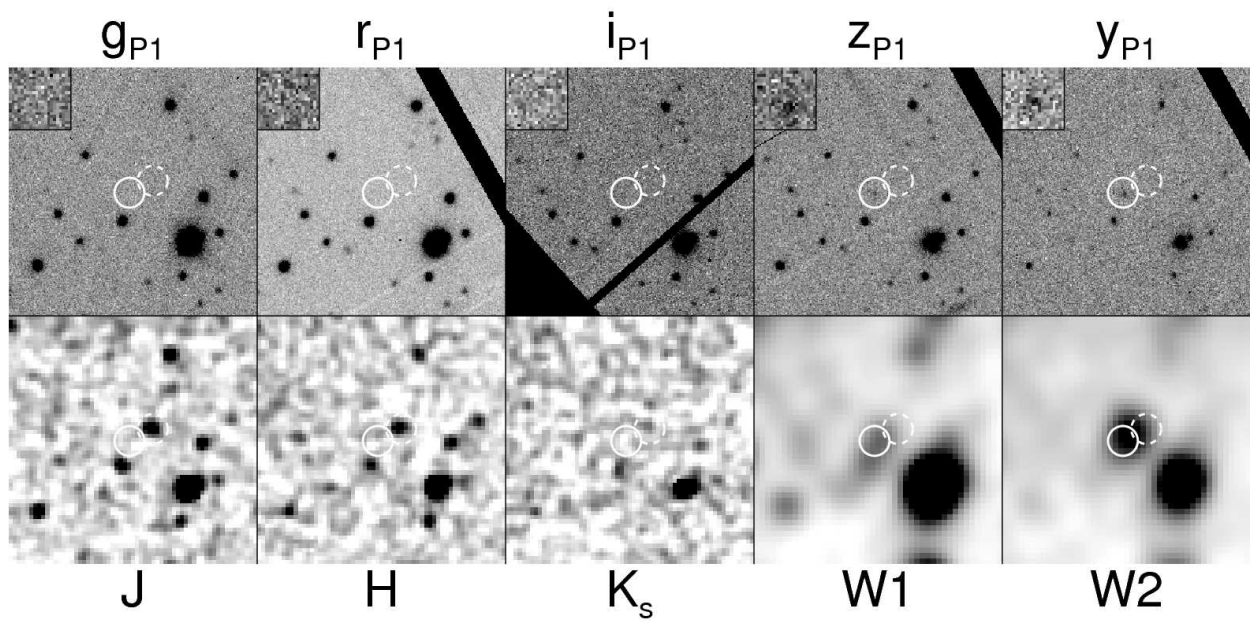}
\caption{Images of HIP 38939B from Pan-STARRS1 ($g_{P1}$, $r_{P1}$, $i_{P1}$, $z_{P1}$, $y_{P1}$), 2MASS ($J$, $H$, $K_s$) and {\it WISE} ($W1$, $W2$). The white circles are centered on the position in the Pan-STARRS1 $y_{p1}$ image (solid line) and the 2MASS position (dashed line). Images are one arcminute across with North up and East left. The cutout images in the upper left of the Pan-STARRS1 panels are 5$\arcsec$ across and centered on the position in the Pan-STARRS1 $y_{P1}$ image shown. \label{stamps}}

\end{center}
\end{figure}

Details of the companion are shown in Table~\ref{details}. Its proper motion is a good match to that of the primary, differing by 1.5$\sigma$. At the distance to the primary (18.5$\pm$0.4 pc; \citealt{vanLeeuwen2007}) the angular separation of 88 arcseconds means a projected separation of 1630 AU. Finder charts are shown in Figure~\ref{stamps}. The relatively dense field is due to the low Galactic latitude of HIP~38939 ($b=1.88^{\circ}$).  This explains why the object has gone undiscovered until now, as most brown dwarf surveys have avoided the Galactic plane.
\begin{deluxetable}{lcc}
\tablecolumns{3}
\tablewidth{0pc}
\tablecaption{\label{details}The HIP 38939 system.}
\tablehead{
&\colhead{HIP 38939A}&\colhead{HIP 38939B}}
\startdata
RA (J2000)&\phs07 58 04.13\tablenotemark{a}\tablenotetext{a}{All astrometry for the primary comes from \cite{vanLeeuwen2007}.}&\phs07 58 01.61\tablenotemark{b}\tablenotetext{b}{Epoch 2010.0}\\
Dec (J2000)&$-$25 37 33.7\phn\tablenotemark{a}&$-$25 39 01.4\phn\tablenotemark{b}\\
$\mu_{\alpha}\cos(\delta)$ ($\arcsec$/yr)&\phs0.362$\pm$0.001\tablenotemark{a}&\phs0.353$\pm$0.006\\
$\mu_{\delta}$ ($\arcsec$/yr)&$-$0.245$\pm$0.001\tablenotemark{a}&$-$0.244$\pm$0.006\\
$\pi$($\arcsec$)&\phs0.054$\pm$0.001\tablenotemark{a}&\\
Spectral Type&K4V\tablenotemark{c}\tablenotetext{c}{\cite{Gray2006}}&T4.5V\tablenotemark{d}\tablenotetext{d}{This work}\\
$B$ (mag)&9.50\tablenotemark{e}\tablenotetext{e}{\cite{Koen2010}}&\\
$V$ (mag)&8.44\tablenotemark{e}&\\
$z_{P1}$  (AB mag)&&20.17$\pm$0.15\tablenotemark{f}\tablenotetext{f}{Pan-STARRS1 Image Processing Pipeline.  Zeropoints are provisionally calibrated relative to the 2MASS \citep{Skrutskie2006}, Tycho \citep{Hog2000} and USNO-B \citep{Monet2003} surveys.}\\
$y_{P1}$ (AB mag)&&18.66$\pm$0.08\tablenotemark{f}\\
2MASS $J$ (mag)&6.51$\pm$0.02\tablenotemark{g}\tablenotetext{g}{\cite{Skrutskie2006} $>$ indicates the 2MASS magnitude is a 97\% confidence upper limit.}&16.12$\pm$0.08\tablenotemark{g}\\
2MASS $H$ (mag)&5.94$\pm$0.02\tablenotemark{g}&15.80$\pm$0.12\tablenotemark{g}\\
2MASS $K_s$ (mag)&5.83$\pm$0.02\tablenotemark{g}&$>$15.86\tablenotemark{g}\\
$M_{\rm bol}$ (mag)&7.93\tablenotemark{h}\tablenotetext{h}{\cite{Casagrande2010}}&17.04$\pm$0.18\tablenotemark{d}\\
$T_{eff}$ (K)&4683$\pm$35\tablenotemark{h}&1090$\pm^{70\tablenotemark{i}\tablenotetext{i}{This work, evolutionary models, see Section \ref{evmodels_sec}.}}_{60}$\\
&&1100\tablenotemark{j}\tablenotetext{j}{This work, atmospheric models, see Section \ref{atmodels_sec}.}\\
$\log_{10} g$ (cm/s$^{2})$&4.5\tablenotemark{h}&5.0$\pm0.3$\tablenotemark{i}\\
&&4.5\tablenotemark{i}\\
\cutinhead{Synthesized colors$^d$}
$J-H$(2MASS) (mags)&&0.139$\pm$0.003\\
$H-K_s$(2MASS) (mags)&&$-$0.104$\pm$0.006\\
$J-K_s$(2MASS) (mags)&&0.03$\pm$0.006\\
$J_{MKO} - J_{2MASS}$ (mags)&&$-$0.217$\pm$0.007\\
$J-H$(MKO) (mags)&&$-$0.127$\pm$0.003\\
$H-K$(MKO) (mags)&&$-$0.191$\pm$0.007\\
$J-K$(MKO) (mags)&&$-$0.318$\pm$0.006\\
\hline
\enddata
\tablewidth{20pc}
\vspace{-0.7cm}
\normalsize
\end{deluxetable}
\subsection{Data from other surveys}
HIP 38939B does not appear in Data Release 8 of the Sloan Digital Sky Survey \citep{SDSS8} and is too far south to appear in the UKIDSS Galactic Plane Survey \citep{LucasGPS}, the IPHAS \citep{Drew2005} or UVEX surveys \citep{Groot2009}. The object falls in the area of the {\it WISE} \citep{Wright2010} Preliminary Data Release but has no detection in the catalogue. However visual inspection indicated there was a bright $W2$-band object at the PS1 position of the source. We retrieved FITS images from the {\it WISE} database and ran SExtractor \citep{SExtractor} on the area. We then used the {\it WISE} catalog magnitudes  for other objects in the field to calculate the zero points of the images. The $W2$ magnitude for this HIP 38939B is $13.82\pm0.04$ mag, with a distance of 0.6$\arcsec$ from the 2010.0 position. In the $W1$-band there is a faint ($W1=15.54\pm0.13$ mag) object 4.8$\arcsec$ away from the PS1 position which appears to be blended with another object of similar magnitude (see Figure~\ref{stamps}). Hence we could not determine an accurate $W1$-band magnitude for HIP 38939B.
\subsection{Likelihood of chance alignment}
We searched for companions around 9363 nearby Hipparcos with projected separations
less than 10,000~AU. The sum of these search areas around our stars comes to 260
sq.deg. The dwarfarchives.org website~\footnote{http://dwarfarchives.org} reports 90 T dwarfs known to an approximate 2MASS depth of $J \sim 16.5$ (about 1 per
400~sq.deg.). Hence there is a reasonable chance that one field T dwarf will appear within our search
area by coincidence. The probability of a random object having the same proper motion as HIP~38939A was estimated
by taking objects from our high-speed Pan-STARRS1-2MASS catalogue within one degree of the primary and
estimating what fraction had a similar proper motion to the primary. Only 5 out of 31349 objects had proper
motions within 5$\sigma$. Hence we believe (assuming the velocity distribution of T dwarfs is not substantially different from that of objects in the HIP 38939 field) that the probability of an unassociated T
dwarf having the same proper motion as the primary is $<$0.02\%.
\section{Spectroscopy}
\begin{figure}[h]
\vskip -2.5in
\begin{center}
\includegraphics[width=5.5in,angle=0]{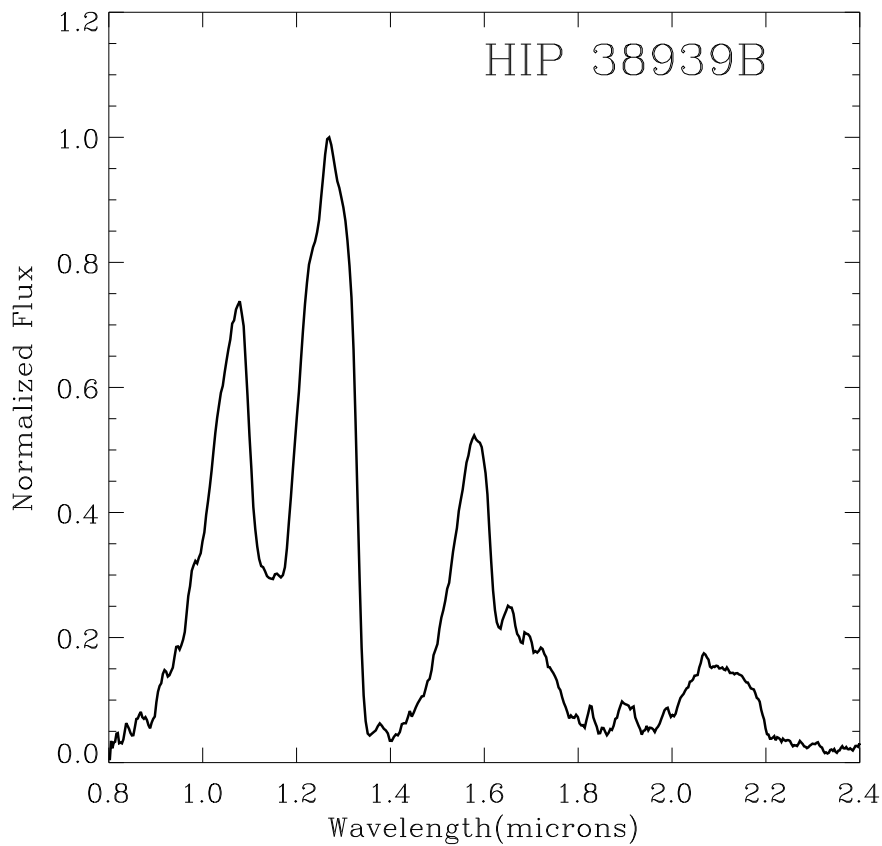}
\vskip -2ex
\caption{The near-infrared spectrum of HIP 38939B. The absorption bands characteristic of a T dwarf can clearly be seen. Note the spectrum has been normalized to its maximum value. \label{spectrum}}

\end{center}
\end{figure}

Spectroscopic confirmation of HIP 38939B was obtained using the SpeX spectrograph \citep{Rayner2003} on the NASA Infrared Telescope Facility on Mauna Kea on 31 March 2011 UT. Observing conditions were clear with good seeing. We used the low resolution, single-order prism mode to obtain a $0.8-2.5$ $\mu$m spectrum. Observations were made at an airmass of 1.89 using the 0.5$\times$15 arcsecond slit, yielding a spectral resolution of $\sim$120. The total integration time was 960 seconds. The slit was oriented at the parallactic angle to minimize atmospheric dispersion. We nodded the telescope in an ABBA pattern with individual exposures of 120s. The spectra were extracted, wavelength calibrated  and telluric corrected using a contemporaneously observed A0V star in the SpeXtool package (\citealt{Cushing2004}, \citealt{Vacca2003}). The final reduced spectrum is in Figure~\ref{spectrum}.

\subsection{Spectral Classification}  
We assigned a near-IR spectral type for HIP 38939B using the flux indices of \cite{Burgasser2006} and the polynomial relations of \cite{Burgasser2007}. The spectrum was also visually inspected by comparing with IRTF/SpeX prism spectral standards from \cite{Burgasser2006}. In this process the spectrum was normalized to the peak fluxes of the standards in the $J$, $H$ and $K$ bands individually and the depth of the H$_2$O and CH$_4$ absorption bands were examined. These spectral indices are shown in Table~\ref{spexclass}. From visual comparison, we find the best fitting spectral type is T5: our spectral index measurements indicate a type of T4.5. We allow our spectral index measurement to take precedent over the visual comparison and we assign a spectral type of T4.5. 

From our spectrum we synthesized a color term between the MKO and 2MASS $J$ bands. This gave a $J$ band magnitude in the MKO system \citep{Tokunaga2002} of 15.90$\pm$0.08 mag. We then used this combined with an average of the bright and faint spectral type to absolute magnitude relations in \cite{Liu2006} to estimate a photometric distance to HIP 38939B of 20.8$^{+2.8}_{-2.1}$ pc. This differs from the Hipparcos distance to the primary of 18.5$\pm$0.4 pc by less than $1\sigma$. Using the similar distance estimator of \cite{Goldman2010} we get a distance of 20.8$\pm$4.3 pc, within $1\sigma$ of the distance to the primary.

\begin{deluxetable}{cccccccc}
\tablecolumns{3}
\tablewidth{0pc}
\tabletypesize{\scriptsize}
\tablecaption{\label{spexclass} Spectral indices for HIP 38939B.}
\tablehead{\colhead{H$_2$O-J}&\colhead{CH$_4$-J}&\colhead{H$_2$O-H}&\colhead{CH$_4$-H}&\colhead{CH$_4$-K}&\colhead{avg/RMS}&\colhead{Visual}&\colhead{Final}}
\startdata
0.303 (T4.5) & 0.492 (T3.6) & 0.388 (T4.4) & 0.453 (T4.9) & 0.244 (T4.8)& T4.4$\pm$0.5&T5&T4.5$\pm$0.5\\
\enddata
\end{deluxetable}
\section{Discussion}
\subsection{The characteristics of HIP 38939A}
\cite{Gray2006} classify HIP 38939A as a K4V. This is consistent with the value of $T_{eff}=4683 \pm 35$ K measured by \cite{Casagrande2010}, who also estimated a sub-solar metallicity of [Fe/H]=-0.24, while \cite{Santos2005} measured 4660 $\pm$ 66 K and [Fe/H]=-0.33 $\pm$ 0.07. The Hipparcos satellite measured a trigonometric parallax of $54.91\pm1.15$  mas and proper motions of $\mu_{\alpha}\cos{\delta}=362.78\pm0.070$  mas/yr and $\mu_{\delta}=-245.89\pm0.63$ mas/yr \citep{vanLeeuwen2007}. 

In order to properly characterize HIP 38939B we must constrain its age by estimating the age of its primary. One of the key estimators for the ages of F, G and K type star is chromospheric activity (\citealt{Wilson1963}, \citealt{Soderblom1991}, \citealt{Donahue1998}, \citealt{Mamajek2008}). This is the by-product of the faster rotation rates of younger stars which drive correspondingly more active dynamos and hence higher chromospheric activity. The chromospheric activity is characterised by the emission in the calcium $H$ and $K$ lines and is typically expressed as,
\begin{equation}
R'_{HK} = R_{HK} - R_{phot}
\label{rhkprime}
\end{equation}
Where $R_{phot}$ is the photospheric contribution to the flux in the Calcium H and K lines, and,
\begin{equation}
R_{HK} = 1.34 \times 10^{-4} C_{cf} S
\label{rhk}
\end{equation}
Here $C_{cf}$ is a correction factor dependent on color (\citealt{Middelkoop1982}, \citealt{Noyes1984}) and $S$ is a ratio of spectral indices in the H and K lines to continuum indices as defined by \cite{Vaughan1978}.

\cite{Gray2006} classify HIP 38939A as being ``active" meaning it has a $\log_{10}R'_{HK}$ value between $-$4.75 and $-$4.2. \cite{Wright2004} measure the mean $S$ value of the object to be 0.757, but due to concerns about calculating the correction factor do not measure an age. We use the relations in \cite{Noyes1984} to calculate that $C_{cf}=0.314$ based on $B-V=1.06$ \citep{Koen2010}. As \cite{Noyes1984} suggest that the photospheric contribution to the Calcium $H$ and $K$ emission for stars redder than $B-V=1.0$ is negligible, we assume $R_{phot}=0$. Hence we derive a $\log_{10}R'_{HK}$ value of $-$4.5.

Comparing to individual cluster datasets from the 680 Myr old \citep{Perryman1998} Hyades \citep{Soderblom1985} and the 500 Myr old Ursa Major moving group \citep{King2003}, we find that HIP 38939A has chromospheric emission consistent with the ages of these two young clusters/associations. However we find that our activity value is not a good fit when compared to data from the younger (100-120 Myr \citealt{Martin1998}) Pleiades \citep{Soderblom1993}.

The most current age-activity-rotation relation is that of \cite{Mamajek2008} which unfortunately does not apply to stars redder than $B-V=0.9$. Using the earlier relation of \cite{Donahue1998} we derive an age of 870 Myr. \cite{Donahue1998} estimates a spread of $0.5$ dex for the age of the Sun from a single measurement of the chromospheric activity at any point across the solar cycle. This value is consistent with the results for younger, coeval binaries and is larger than the intrinsic variability of 13\% quoted by \cite{Wright2004} for their $S$ values. Hence we assign errors to the Donahue value of 900$\pm^{1900}_{600}$ Myr. 

Another indicator of stellar activity and hence youth is X-ray emission. We searched the ROSAT Faint Source Catalogue \citep{Voges2000} and identified an extremely faint source 18$\arcsec$ away from HIP 38939A. This distance is slightly larger than the quoted ROSAT $1\sigma$ positional uncertainty for this source of 15$\arcsec$. As there is always the possibility of association with an unrelated source, we carried out the following calculation. There are 25 ROSAT sources within 3 degrees of the primary. This corresponds to a surface density of 0.88/sq.deg. If we assume a ROSAT source has to be within 30" of the primary to be associated then the probability of a chance alignment is 0.02\%. Hence we assume the X-ray source is associated with HIP 38939A. The source has X-ray counts of $(3.66\pm1.48)\times10^{-2}$ s$^{-1}$ and a hardness ratio of $HR1=-0.76\pm0.27$. Using the relations of \cite{Schmitt1995} we calculate an X-ray luminosity of $L_X=(6.43\pm3.37)\times10^{27}$ ergs/s. \cite{Casagrande2006} derive a bolometric luminosity of 0.175 $L_{\odot}$ for HIP 38939A. From this we calculate a fractional X-ray luminosity $R_X=\log(L_{X}/L_{\rm bol})=\log((9.55\pm5.00)\times10^{-6})$. Using the X-ray luminosity - age relation from \cite{Mamajek2008}, we derive an age of $900\pm_{600}^{1600}$ Myr taking in to account both the measurement errors on the X-ray flux and the quoted 0.4 dex uncertainty in the age relation. While this is a low significance X-ray detection, we note that the X-ray derived age is in agreement with those derived from calcium H and K emission.

\subsection{Comparison of HIP 38939B with evolutionary models}
\label{evmodels_sec}
\begin{figure}[htbp]
\begin{center}
\includegraphics[scale=.70,angle=90]{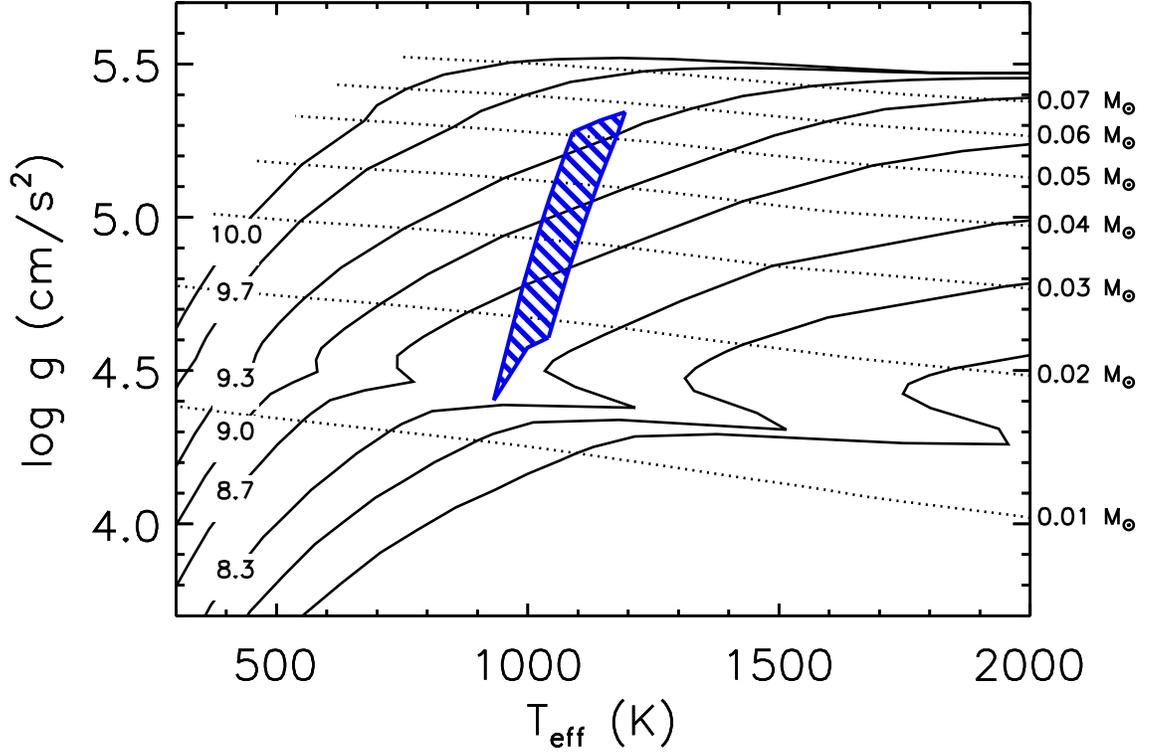}
\caption{Our derived age and bolometric luminosity for HIP 38939B plotted against the evolutionary models of \cite{Burrows1997}. The blue-colored hatched areas indicate the constraints on the parameters given the uncertainties on $M_{\rm bol}$ and the age of the primary. The solid lines represent isochrones with the annotated values of $\log$(age). The two unmarked isochrones at the bottom are $\log$(age)=8.0 and 7.7. The dashed lines are lines of equal mass with the appropriate values noted on the right-hand side of the plot. See Table~\ref{sec_prop} for more details on the individual masses for each particular age and the Monte Carlo simulated parameters. \label{evmodels}}

\end{center}
\end{figure}

We derived the bolometric luminosity for HIP 38939B by converting the 2MASS $J$-band magnitude into the MKO system using a color term synthesized from our spectrum, converting to absolute magnitude using the Hipparcos parallax for the primary and applying the bolometric correction relation found in \cite{Liu2010}. This resulted in a bolometric magnitude of $M_{\rm bol}=17.04\pm0.18$ mag. We then compared the range of ages derived for the primary along and our calculated $M_{\rm bol}$ to the evolutionary models of \cite{Burrows1997}. Figure~\ref{evmodels} shows where the secondary lies in comparison with these evolutionary models. Table~\ref{sec_prop} also shows the values of various parameters for different ages. Additionally we ran a Monte Carlo simulation with our calculated bolometric magnitude and with a uniformly distributed range of ages derived from the primary. The results for these are also shown in Table~\ref{sec_prop}.

\begin{deluxetable}{cccc}
\tablecolumns{3}
\tablewidth{0pc}
\tablecaption{\label{sec_prop} Derived properties for HIP 38939B using our inferred age range for the primary.}
\tablehead{\colhead{$\log_{10}$(age)}&\colhead{Mass}&\colhead{$T_{eff}$}&\colhead{$\log(g)$}\\
\colhead{yr}&\colhead{$M_{\rm Jup}$}&\colhead{K}&\colhead{cm/s$^{2}$}}
\startdata
8.40&19&1020&4.6\\
8.70&27&1060.&4.8\\
8.90&34&1090&5.0\\
8.95&36&1090&5.0\\
9.00&38&1100&5.0\\
9.05&40&1110&5.1\\
9.10&42&1120&5.1\\
9.30&52&1150&5.3\\
9.40&59&1170&5.3\\
\cutinhead{Monte Carlo Results}
8.90$\pm$0.5&34$_{-11}^{+16}$&1090$_{-60}^{+70}$&5.0$\pm0.3$\\
\enddata
\end{deluxetable}
\subsection{Comparison with atmosphere models}
\label{atmodels_sec}
Atmospheric models provide an independent means of deriving the physical properties of HIP~38989B.
We fit the solar metallicity BT-Settl-2010 models \citep{Allard2010} to our SpeX spectrum 
following the procedures in \cite{Cushing2008} and \cite{Bowler2009}.  The grid of models spans 
effective temperatures between
500--1500~K ($\Delta T_\mathrm{eff}$=100~K) and gravities between 10$^{4.0}$--10$^{5.5}$ 
(cgs; $\Delta$log~$g$=0.5).  The SpeX spectrum was first flux calibrated to the 2MASS
$J$-band photometry and then fit between 0.8--2.4~$\mu$m in a Monte Carlo fashion.  
The 1.60--1.65~$\mu$m region was excluded from the fits because the methane line list in the
models is incomplete at those wavelengths (e.g., Saumon et al. 2007).
For each Monte Carlo trial we alter the flux calibration scaling factor and SpeX spectrum by drawing
random values from normal distributions based on photometric and spectral measurement
uncertainties, respectively.  We then fit the grid of models to each artificial spectrum and repeat the process 10$^3$ times.

\begin{figure}[htbp]
\vskip -1in
\hskip -0.8in
\includegraphics[scale=.9]{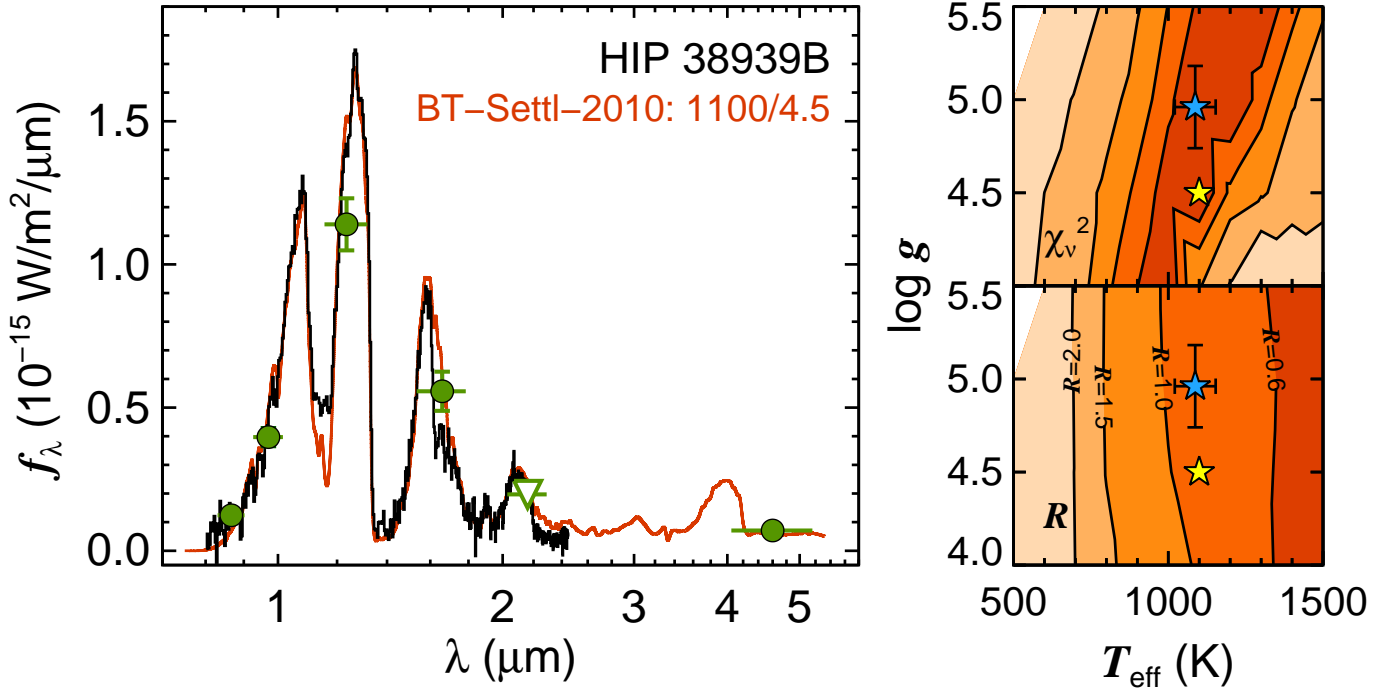}
\vskip -12ex
\caption{{\it Left:} The comparison between the observed SpeX spectrum (black) and photometry (green circles for data points, green triangles for upper limits) and the best fitting model atmosphere (red) from the BT-Settl-2010 models \citep{Allard2010}. The spectrum is flux calibrated using the $J$-band photometry from 2MASS. The magnitude to flux conversions used the zeropoints and effective wavelengths from \cite{Rieke2008} and \cite{Wright2010} for 2MASS and {\it WISE} respectively, and the PS1 image processing pipeline AB zeropoints and the effective wavelengths from Tonry (in prep). for the PS1 data.  The horizontal error bars represent the filter bandpass. The {\it WISE} 4.6~$\mu$m point is consistent with the best-fit model atmosphere, which assumes chemical equilibrium (no vertical mixing). {\it Upper right:} The $\chi_{\nu}^2$ contours for the comparison between the models and the observed spectrum, the contours represent values of 70, 100, 150, and 250. The best fitting model (1100 K, $\log g$=4.5) is shown as a yellow star with the blue star and error bars representing the results of Monte Carlo calculations based on the evolutionary models of \cite{Burrows1997}. The two agree within the stated error in $T_{eff}$ and differ by only one grid-point in $\log g$. When the region of the spectrum from 1.6-1.65 \micron (previously excluded due to incomplete model line lists) is included, the best fitting gravity moves to $\log g$=5.5. This is still within one grid point of the evolutionary model calculations. {\it Lower right:} The different radii as a fraction of the radius of Jupiter from the model atmosphere fitting.\label{atmodels}. See \cite{Bowler2009} for details on this method.}
\end{figure}

The best-fitting BT-Settl-2010 model has $T_\mathrm{eff}$=1100~K and $\log$~$g$=4.5\footnote{We also carried out a fit to the models including the 1.6-1.65 $\mu$m spectral region. The best fit model was $T_\mathrm{eff}$=1200~K/$\log$~$g$=5.5, which differs slightly from the 1100~K/4.5~dex best fit when that wavelength region is excluded.}.  This is in 
good agreement with the evolutionary model predictions of $T_\mathrm{eff}$=1090$^{+70}_{-60}$~K 
and log~$g$=5.0$\pm0.3$ especially considering the coarse gridding of the atmospheric models. In Figure~\ref{atmodels}
we show the flux-calibrated SpeX spectrum along with photometry from PS1 ($y_{P1}$ and $z_{P1}$ bands),
2MASS ($J$, $H$, and an upper limit on $K_S$), and {\it WISE} ($W2$).  
The monochromatic flux densities agree well with the spectrum between 0.8--5~$\mu$m.  The best-fitting
atmospheric model provides a good match to the spectrum except at $\sim$1.15~$\mu$m, 1.65~$\mu$m,
and 2.15$\mu$m, which correspond to absorption features from CH$_4$. 

Using our SpeX spectrum together with the 
$J$-band photometry measured by 2MASS, we also synthesize photometry of HIP~38989B in the 2MASS $H$ and $K_s$ bands and 
our spectral measurement errors.  The synthesized magnitudes for 10$^4$ Monte Carlo trials
are $H$=15.98$\pm$0.08,
and $K_S$=16.09$\pm$0.08 mag.  The $H$ band synthetic magnitude is within 2$\sigma$ of the 
2MASS measurement, which is near the detection limit of the survey.  The synthesized
colors are $J-H$=0.139$\pm$0.003, $H-K_S$=--0.104$\pm$0.006 and $J-K_S$=0.03$\pm$0.006 mag.
The uncertainties in the synthesized colors are much smaller than the magnitudes because they
only include spectral measurement uncertainties.

\subsection{The population of benchmark brown dwarfs}
\begin{figure}[htbp]
\begin{center}
\includegraphics[scale=.75]{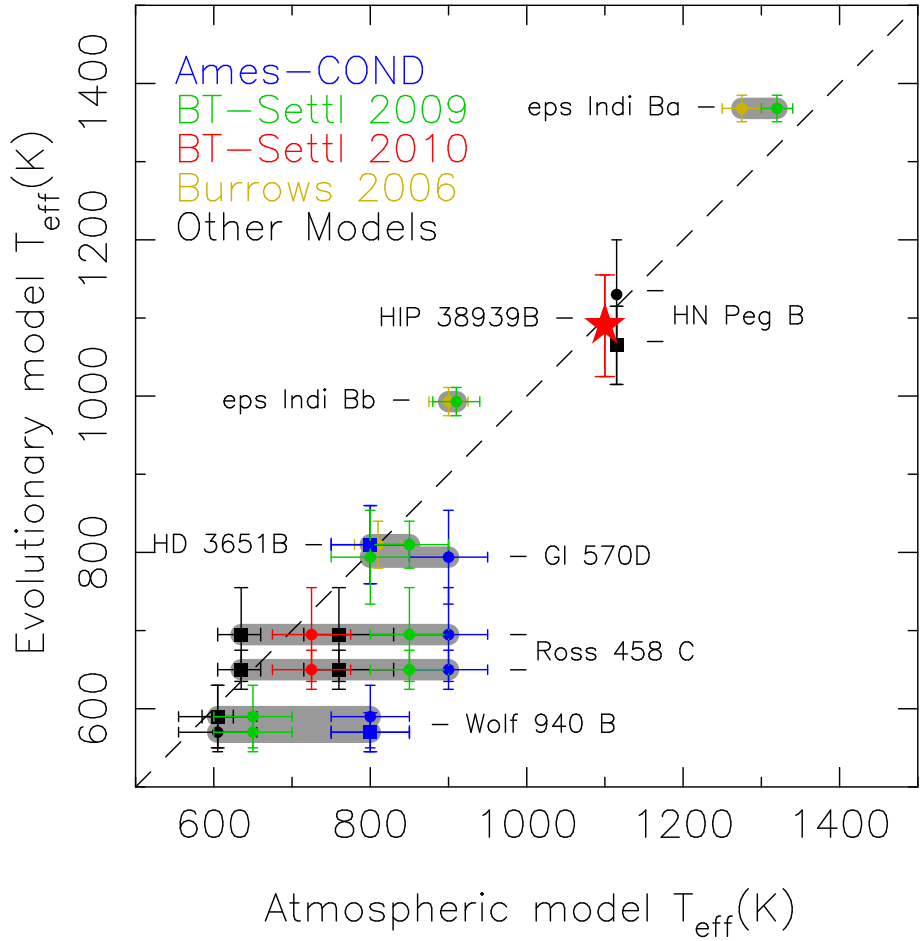}
\caption{\label{evatfig} The comparison of the evolutionary and atmospheric model effective temperatures for the objects listed in Table~\ref{model_comp}. The error bars represent the range of values quoted in the references and the dashed line marks a one to one agreement between the two temperatures. The gray bars link the datapoints of objects with multiple values of the atmospheric effective temperature. HIP 38939B is marked by a star. The ``other models" plotted are model calculations based on \cite{Saumon2008} carried out for \cite{Leggett2010}, \cite{Leggett2008} and \cite{Burgasser2010a}. Square symbols represent objects where the evolutionary model effective temperature is derived from a model which uses the same  atmospheric model as a boundary condidtion as was used to derive the atmospheric model effective temperature. It appears that for the coolest T dwarfs, only the most recent models provide good agreement between the two temperatures. For the components of $\epsilon$ Indi B the discrepancy between the two temperatures may be due to an incorrect age for the system (\citealt{King2010}, \citealt{Liu2010}).}

\end{center}
\end{figure}

As noted in Section 1, brown dwarf binaries can be used to test atmospheric and evolutionary models of substellar objects. Atmospheric models of brown dwarfs will not fully match the shape of observed spectra due to missing opacity lists for H$_2$O, CH$_4$ and NH$_3$ \citep{Leggett2007}. In order to produce a more complete picture of how these models compare with the growing population of benchmark brown dwarfs, we compiled a list of all T dwarf companions with effective temperatures derived from both model atmospheres and evolutionary models. In order to be included, an object had to have an effective temperature derived from its bolometric luminosity and a separate temperature derived from model atmosphere fits to the observed spectrum. These are listed in Table~\ref{model_comp} along with the models used for each calculation. The two temperatures for each object are plotted against each other in Figure~\ref{evatfig}. Evolutionary models depend on atmospheric models to set their boundary conditions. In many of the comparisons we make, the evolutionary model boundary atmosphere is different from that used to derive the atmospheric effective temperature. In cases where the same atmospheric model is used this has been noted in both Figure~\ref{evatfig} and Table~\ref{model_comp}. Many of our objects have evolutionary model effective temperatures derived from different sources. Table~\ref{model_comp} shows that there is no significant difference between the derived temperatures. \cite{Luhman2007} use \cite{Burrows1997} and \cite{Baraffe2003} evolutionary models to derive effective temperatures for HN Peg B and HD 3651 B finding that there is no significant difference between the two. There appears to be a bias towards lower temperature objects having atmospheric temperatures which are higher than those derived from evolutionary models. It should be noted that the best fits for these objects are those from the most recent models, BT-Settl 2010 \citep{Allard2010} for Ross 458C \cite{Burningham2011}, fits to \cite{Saumon2008} by \cite{Burgasser2010a} and a set of models calculated by Saumon for \cite{Leggett2010}'s study of Wolf 940B. \cite{Dupuy2009a} noted that atmosphere models for late-M dwarfs also predict higher temperatures than the evolutionary models.
\begin{landscape}
\begin{deluxetable}{lc|cll|cll}
\tablecolumns{3}
\tablewidth{0pc}
\tabletypesize{\scriptsize}
\tablecaption{\label{model_comp} Known T dwarf companions to main sequence stars with effective temperatures from model atmospheres ($T_{eff,atm}$) and evolutionary models ($T_{eff,evol}$).}
\tablehead{\colhead{Object}&\colhead{SpT}&\colhead{$T_{eff,evol}$}&\colhead{Evolutionary Model}&\colhead{Reference}&\colhead{$T_{eff,atm}$}&\colhead{Atmosphere Model}&\colhead{Reference}\\
&\colhead{K}&&&\colhead{K}&&}
\startdata
HIP 38939B&T4.5\tablenotemark{a}&1090$\pm^{70}_{60}$&\cite{Burrows1997}&This work&1100&\cite{Allard2010}&this work\\
$\epsilon$ Indi Ba&T1\tablenotemark{b}&1368$\pm$17\tablenotemark{c}&\cite{Baraffe2003}&\cite{King2010}&1320$\pm$20&\cite{Allard2010}&\cite{King2010}\\
&&&&&1275$\pm$25&\cite{Burrows2006}&\cite{Kasper2009}\\
$\epsilon$ Indi Bb&T6\tablenotemark{b}&993$\pm$18\tablenotemark{c}&\cite{Baraffe2003}&\cite{King2010}&910$\pm$30&\cite{Allard2010}&\cite{King2010}\\
&&&&&900$\pm$25&\cite{Burrows2006}&\cite{Kasper2009}\\
HN Peg B&T2.5\tablenotemark{d}&1065$\pm$50&\cite{Leggett2008}&\cite{Leggett2008}&1115&\cite{Leggett2008}&\cite{Leggett2008}*\\
&&1130$\pm$70&\cite{Burrows1997} + \cite{Baraffe2003}&\cite{Luhman2007}&&&\\
Gl 570 D&T7.5\tablenotemark{e}&794$\pm$60&\cite{Burrows1997}&\cite{Geballe2001}&800$\pm$20&\cite{Burrows2006}&\cite{Burgasser2006b}\\
&&&&&900$\pm$50&\cite{Allard2001}&\cite{Liu2011a}\\
&&&&&800$\pm$50&\cite{Allard2010a}&\cite{Liu2011a}\\
HD 3651 B&T7.5\tablenotemark{d}&810$\pm30$&\cite{Burrows1997}&\cite{Liu2007a}&810$\pm$30&\cite{Burrows2006}&\cite{Burgasser2007b}\\
&&810$\pm$50$\pm$70&\cite{Burrows1997} + \cite{Baraffe2003}&\cite{Luhman2007}&800$\pm$50&\cite{Allard2001}&\cite{Liu2011a}*\\
&&&&&850$\pm$50&\cite{Allard2010a}&\cite{Liu2011a}\\
Ross 458 C&T8.5p\tablenotemark{f}&695$\pm$60&\cite{Saumon2008}&\cite{Burningham2011}&725$\pm$25&\cite{Allard2010}&\cite{Burningham2011}\\
&&650$\pm$25&\cite{Saumon2008}&\cite{Burgasser2010a}&635$\pm^{25}_{30}$&\cite{Saumon2008}&\cite{Burgasser2010a}\tablenotemark{h}*\\
&&&&&760$\pm^{70}_{45}$&\cite{Saumon2008}&\cite{Burgasser2010a}\tablenotemark{i}*\\
&&&&&900$\pm$50&\cite{Allard2001}&\cite{Liu2011a}\\
&&&&&850$\pm$50&\cite{Allard2010a}&\cite{Liu2011a}\\
Wolf 940 B&T8.5\tablenotemark{i}&590$\pm$40&\cite{Saumon2008}&\cite{Leggett2010}&605$\pm$20&\cite{Leggett2010}&\cite{Leggett2010}*\\
&&570$\pm$25&\cite{Baraffe2003}&\cite{Burningham2009}&800$\pm$50&\cite{Allard2001}&\cite{Liu2011a}*\\
&&&&&650$\pm$50&\cite{Allard2010a}&\cite{Liu2011a}\\
\enddata
\tablenotetext{*}{Evolutionary model boundary atmosphere model is the same as the model used to derived the atmospheric effective temperature.}
\tablenotetext{a}{This work}
\tablenotetext{b}{\cite{McCaughrean2004}}
\tablenotetext{c}{This is based on an age of 4.0$\pm$0.3. A number of different indicators give a wide range of ages for the $\epsilon$ Indi system. An older age would bring the evolutionary models into better agreement with the atmosphere models (\citealt{King2010}, \citealt{Liu2010}}
\tablenotetext{d}{\cite{Luhman2007}}
\tablenotetext{e}{\cite{Burgasser2006}}
\tablenotetext{f}{\cite{Burningham2011}}
\tablenotetext{g}{cloudy model}
\tablenotetext{h}{cloudless model}
\tablenotetext{i}{\cite{Burningham2009}}
\end{deluxetable} 
\end{landscape}
\section{Conclusions}
We have discovered a T4.5 companion to the nearby K4V star HIP 38939 using Pan-STARRS1 and 2MASS data. We estimate the age of the primary as 900$\pm^{1900}_{600}$ Myr from which we deduce that the companion has a mass in the range 38$\pm$20 $M_{Jup}$. The effective temperatures and surface gravities derived from model atmosphere fitting and evolutionary models are consistent. Given its close proximity to Earth (18.4 pc), this object will be a high priority target to identify if it itself is a binary.  Such objects are of great importance for testing physical models of brown dwarfs~\citep{Liu2008}. This discovery indicates Pan-STARRS1 data can be used to explore crowded regions of the sky such as the Galactic plane often ignored by other studies. In the future the wide-field, multi-epoch nature of the Pan-STARRS1 3$\pi$ survey will lead to a well-defined sample of wide substellar companions to nearby main sequence stars. 

\acknowledgements
The PS1 Surveys have been made possible through contributions of the Institute for Astronomy, the University of Hawaii, the Pan-STARRS Project Office, the Max-Planck Society and its participating institutes, the Max Planck Institute for Astronomy, Heidelberg and the Max Planck Institute for Extraterrestrial Physics, Garching, The Johns Hopkins University, the University of Durham, the University of Edinburgh, Queen's University Belfast, the Harvard-Smithsonian Center for Astrophysics, and the Los Cumbres Observatory Global Telescope Network, Incorporated, the National Central University of Taiwan, and the National Aeronautics and Space Administration under Grant No. NNX08AR22G issued through the Planetary Science Division of the NASA Science Mission Directorate.They would like to thank Dave Griep for assisting with the IRTF observations. This research has benefitted from the SpeX Prism Spectral Libraries, maintained by Adam Burgasser at http://www.browndwarfs.org/spexprism. This publication makes use of data products from the Two Micron All Sky Survey, which is a joint project of the University of Massachusetts and the Infrared Processing and Analysis Center/California Institute of Technology, funded by the National Aeronautics and Space Administration and the National Science Foundation. This research has benefitted from the M, L, and T dwarf compendium housed at DwarfArchives.org and maintained by Chris Gelino, Davy Kirkpatrick, and Adam Burgasser. This research was supported by NSF grants AST-0507833 and AST09-09222 (awarded to MCL), AST-0709460 (awarded to EAM), AFRL Cooperative Agreement FA9451-06-2-0338, and DFG-
Sonderforschungsbereich 881 The Milky Way. Finally, the authors wish to recognize the very
significant cultural role that the summit of Mauna Kea has always had within the indigenous
Hawaiian community. We are most fortunate to conduct observations from this mountain.\\
{\it Facilities:} \facility{IRTF (SpeX)}, \facility{PS1}


\begin{thebibliography}{102}
\expandafter\ifx\csname natexlab\endcsname\relax\def\natexlab#1{#1}\fi

\bibitem[{Aihara {et~al.}(2011)Aihara, {Allende Prieto}, An, Anderson, Aubourg,
  Balbinot, Beers, Berlind, Bickerton, Bizyaev, Blanton, Bochanski, Bolton,
  Bovy, Brandt, Brinkmann, Brown, Brownstein, Busca, Campbell, Carr, Chen,
  Chiappini, Comparat, Connolly, Cortes, Croft, Cuesta, da~Costa, Davenport,
  Dawson, Dhital, Ealet, Ebelke, Edmondson, Eisenstein, Escoffier, Esposito,
  Evans, Fan, {Femen\'{\i}a Castell\'{a}}, Font-Ribera, Frinchaboy, Ge,
  Gillespie, Gilmore, {Gonz\'{a}lez Hern\'{a}ndez}, Gott, Gould, Grebel, Gunn,
  Hamilton, Harding, Harris, Hawley, Hearty, Ho, Hogg, Holtzman, Honscheid,
  Inada, Ivans, Jiang, Johnson, Jordan, Jordan, Kazin, Kirkby, Klaene, Knapp,
  Kneib, Kochanek, Koesterke, Kollmeier, Kron, Lampeitl, Lang, {Le Goff}, Lee,
  Lin, Long, Loomis, Lucatello, Lundgren, Lupton, Ma, MacDonald, Mahadevan,
  Maia, Makler, Malanushenko, Malanushenko, Mandelbaum, Maraston, Margala,
  Masters, McBride, McGehee, McGreer, M\'{e}nard, Miralda-Escud\'{e}, Morrison,
  Mullally, Muna, Munn, Murayama, Myers, Naugle, {Fausti Neto}, Nguyen, Nichol,
  O'Connell, Ogando, Olmstead, Oravetz, Padmanabhan, Palanque-Delabrouille,
  Pan, Pandey, P\^{a}ris, Percival, Petitjean, Pfaffenberger, Pforr, Phleps,
  Pichon, Pieri, Prada, Price-Whelan, Raddick, Ramos, Reyl\'{e}, Rich,
  Richards, Rix, Robin, Rocha-Pinto, Rockosi, Roe, Rollinde, Ross, Ross,
  Rossetto, S\'{a}nchez, Sayres, Schlegel, Schlesinger, Schmidt, Schneider,
  Sheldon, Shu, Simmerer, Simmons, Sivarani, Snedden, Sobeck, Steinmetz,
  Strauss, Szalay, Tanaka, Thakar, Thomas, Tinker, Tofflemire, Tojeiro,
  Tremonti, Vandenberg, {Vargas Maga\~{n}a}, Verde, Vogt, Wake, Wang, Weaver,
  Weinberg, White, White, Yanny, Yasuda, Yeche, \& Zehavi}]{SDSS8}
Aihara, H., {et~al.} 2011, The Astrophysical Journal Supplement Series, 193, 29

\bibitem[{Albert {et~al.}(2011)Albert, Artigau, Delorme, Reyl\'{e}, Forveille,
  Delfosse, \& Willott}]{Albert2011}
Albert, L., Artigau, E., Delorme, P., Reyl\'{e}, C., Forveille, T., Delfosse,
  X., \& Willott, C.~J. 2011, The Astronomical Journal, 141, 203

\bibitem[{Allard \& Freytag(2010)}]{Allard2010a}
Allard, F., \& Freytag, B. 2010, Proceedings of the International Astronomical
  Union, 5, 756

\bibitem[{Allard {et~al.}(2001)Allard, Hauschildt, Alexander, Tamanai, \&
  Schweitzer}]{Allard2001}
Allard, F., Hauschildt, P.~H., Alexander, D.~R., Tamanai, A., \& Schweitzer, A.
  2001, The Astrophysical Journal, 556, 357

\bibitem[{Allard {et~al.}(2010)Allard, Homeier, \& Freytag}]{Allard2010}
Allard, F., Homeier, D., \& Freytag, B. 2010

\bibitem[{Baraffe {et~al.}(2003)Baraffe, Chabrier, Barman, Allard, \&
  Hauschildt}]{Baraffe2003}
Baraffe, I., Chabrier, G., Barman, T.~S., Allard, F., \& Hauschildt, P.~H.
  2003, Astronomy \& Astrophysics, 712, 701

\bibitem[{Beaumont \& Magnier(2010)}]{Beaumont2010}
Beaumont, C.~N., \& Magnier, E.~A. 2010, Publications of the Astronomical
  Society of the Pacific, 122, 1389

\bibitem[{Bertin \& Arnouts(1996)}]{SExtractor}
Bertin, E., \& Arnouts, S. 1996, Astronomy and Astrophysics Supplement, 117,
  393

\bibitem[{Biller {et~al.}(2006)Biller, Kasper, Close, Brandner, \&
  Kellner}]{Biller2006}
Biller, B.~A., Kasper, M., Close, L.~M., Brandner, W., \& Kellner, S. 2006, The
  Astrophysical Journal, 641, L141

\bibitem[{Botticella {et~al.}(2010)Botticella, Trundle, Pastorello, Rodney,
  Rest, Gezari, Smartt, Narayan, Huber, Tonry, Young, Smith, Bresolin, Valenti,
  Kotak, Mattila, Kankare, Wood-Vasey, Riess, Neill, Forster, Martin, Stubbs,
  Burgett, Chambers, Dombeck, Flewelling, Grav, Heasley, Hodapp, Kaiser,
  Kudritzki, Luppino, Lupton, Magnier, Monet, Morgan, Onaka, Price, Rhoads,
  Siegmund, Sweeney, Wainscoat, Waters, Waterson, \&
  Wynn-Williams}]{Botticella2010}
Botticella, M.~T., {et~al.} 2010, The Astrophysical Journal, 717, L52

\bibitem[{Bouvier {et~al.}(2008)Bouvier, Kendall, Meeus, Testi, Moraux,
  Stauffer, James, Cuillandre, Irwin, McCaughrean, Baraffe, \&
  Bertin}]{Bouvier2008}
Bouvier, J., {et~al.} 2008, Astronomy and Astrophysics, 481, 661

\bibitem[{Bowler {et~al.}(2009)Bowler, Liu, \& Cushing}]{Bowler2009}
Bowler, B.~P., Liu, M.~C., \& Cushing, M.~C. 2009, The Astrophysical Journal,
  706, 1114

\bibitem[{Burgasser(2007{\natexlab{a}})}]{Burgasser2007}
Burgasser, A.~J. 2007{\natexlab{a}}, The Astrophysical Journal, 659, 655

\bibitem[{Burgasser(2007{\natexlab{b}})}]{Burgasser2007b}
---. 2007{\natexlab{b}}, The Astrophysical Journal, 658, 617

\bibitem[{Burgasser {et~al.}(2006{\natexlab{a}})Burgasser, Burrows, \&
  Kirkpatrick}]{Burgasser2006b}
Burgasser, A.~J., Burrows, A., \& Kirkpatrick, J.~D. 2006{\natexlab{a}}, The
  Astrophysical Journal, 639, 1095

\bibitem[{Burgasser {et~al.}(2006{\natexlab{b}})Burgasser, Geballe, Leggett,
  Kirkpatrick, \& Golimowski}]{Burgasser2006}
Burgasser, A.~J., Geballe, T.~R., Leggett, S.~K., Kirkpatrick, J.~D., \&
  Golimowski, D.~A. 2006{\natexlab{b}}, The Astrophysical Journal, 637, 1067

\bibitem[{Burgasser {et~al.}(2005)Burgasser, Kirkpatrick, \&
  Lowrance}]{Burgasser2005}
Burgasser, A.~J., Kirkpatrick, J.~D., \& Lowrance, P.~J. 2005, The Astronomical
  Journal, 129, 2849

\bibitem[{Burgasser {et~al.}(2000)Burgasser, Kirkpatrick, Cutri, McCallon,
  Kopan, Gizis, Liebert, Reid, Brown, Monet, Dahn, Beichman, \&
  Skrutskie}]{Burgasser2000}
Burgasser, A.~J., {et~al.} 2000, The Astrophysical Journal, 531, L57

\bibitem[{Burgasser {et~al.}(2010)Burgasser, Simcoe, Bochanski, Saumon,
  Mamajek, Cushing, Marley, Mcmurtry, Pipher, \& Forrest}]{Burgasser2010a}
---. 2010, ApJ, 725, 1405

\bibitem[{Burningham {et~al.}(2009)Burningham, Pinfield, Leggett, Tinney, Liu,
  Homeier, West, Day-Jones, Huelamo, Dupuy, Zhang, Murray, Lodieu, {Barrado y
  Navascu\'{e}s}, Folkes, Galvez-Ortiz, Jones, Lucas, Calderon, \&
  Tamura}]{Burningham2009}
Burningham, B., {et~al.} 2009, Monthly Notices of the Royal Astronomical
  Society, 395, 1237

\bibitem[{Burningham {et~al.}(2011)Burningham, Leggett, Homeier, Saumon, Lucas,
  Pinfield, Tinney, Allard, Marley, Jones, Murray, Ishii, Day-Jones, Gomes, \&
  Zhang}]{Burningham2011}
---. 2011, Monthly Notices of the Royal Astronomical Society, 414, 3590

\bibitem[{Burrows {et~al.}(2006)Burrows, Sudarsky, \& Hubeny}]{Burrows2006}
Burrows, A., Sudarsky, D., \& Hubeny, I. 2006, The Astrophysical Journal, 640,
  1063

\bibitem[{Burrows {et~al.}(1997)Burrows, Marley, Hubbard, Lunine, Guillot,
  Saumon, Freedman, Sudarsky, \& Sharp}]{Burrows1997}
Burrows, A., {et~al.} 1997, The Astrophysical Journal, 491, 856

\bibitem[{Casagrande {et~al.}(2006)Casagrande, Portinari, \&
  Flynn}]{Casagrande2006}
Casagrande, L., Portinari, L., \& Flynn, C. 2006, Monthly Notices of the Royal
  Astronomical Society, 373, 13

\bibitem[{Casagrande {et~al.}(2010)Casagrande, Ram\'{\i}rez, Mel\'{e}ndez,
  Bessell, \& Asplund}]{Casagrande2010}
Casagrande, L., Ram\'{\i}rez, I., Mel\'{e}ndez, J., Bessell, M., \& Asplund, M.
  2010, Astronomy and Astrophysics, 512, A54

\bibitem[{Cushing {et~al.}(2004)Cushing, Vacca, \& Rayner}]{Cushing2004}
Cushing, M.~C., Vacca, W.~D., \& Rayner, J.~T. 2004, Publications of the
  Astronomical Society of the Pacific, 116, 362

\bibitem[{Cushing {et~al.}(2008)Cushing, Marley, Saumon, Kelly, Vacca, Rayner,
  Freedman, Lodders, \& Roellig}]{Cushing2008}
Cushing, M.~C., {et~al.} 2008, The Astrophysical Journal, 678, 1372

\bibitem[{Day-Jones {et~al.}(2008)Day-Jones, Pinfield, Napiwotzki, Burningham,
  Jenkins, Jones, Folkes, Weights, \& Clarke}]{Day-Jones2008}
Day-Jones, A.~C., {et~al.} 2008, Monthly Notices of the Royal Astronomical
  Society, 388, 838

\bibitem[{Day-Jones {et~al.}(2011)Day-Jones, Pinfield, Ruiz, Beaumont,
  Burningham, Gallardo, Gianninas, Bergeron, Napiwotzki, Jenkins, Zhang,
  Murray, Catal\'{a}n, \& Gomes}]{Day-Jones2010}
Day-Jones, a.~C., {et~al.} 2011, Monthly Notices of the Royal Astronomical
  Society, 410, 705

\bibitem[{Deacon {et~al.}(2011)Deacon, Liu, Magnier, Bowler, Goldman, Redstone,
  Burgett, Chambers, Flewelling, Kaiser, Lupton, Morgan, Price, Sweeney, Tonry,
  Wainscoat, \& Waters}]{Deacon2011}
Deacon, N.~R., {et~al.} 2011, The Astronomical Journal, 142, 77

\bibitem[{Donahue(1998)}]{Donahue1998}
Donahue, R.~A. 1998, in ASP Conf. Ser. 154: The Tenth Cambridge Workshop on
  Cool Stars, Stellar Systems and the Sun, ed. R.~Donahue \& J.~Bookbinder,
  1235

\bibitem[{Drew {et~al.}(2005)Drew, Greimel, Irwin, \& Collaboration}]{Drew2005}
Drew, J.~E., Greimel, R., Irwin, M.~J., \& Collaboration, I. 2005, Monthly
  Notices of the Royal Astronomical Society, 362, 25

\bibitem[{Dupuy \& Liu(2011)}]{Dupuy2011}
Dupuy, T.~J., \& Liu, M.~C. 2011, The Astrophysical Journal, 733, 122

\bibitem[{Dupuy {et~al.}(2010)Dupuy, Liu, Bowler, Cushing, Helling, Witte, \&
  Hauschildt}]{Dupuy2009a}
Dupuy, T.~J., Liu, M.~C., Bowler, B.~P., Cushing, M.~C., Helling, C., Witte,
  S., \& Hauschildt, P. 2010, The Astrophysical Journal, 721, 1725

\bibitem[{Dupuy {et~al.}(2009)Dupuy, Liu, \& Ireland}]{Dupuy2008}
Dupuy, T.~J., Liu, M.~C., \& Ireland, M.~J. 2009, The Astrophysical Journal,
  692, 729

\bibitem[{Faherty {et~al.}(2010)Faherty, Burgasser, West, Bochanski, Cruz,
  Shara, \& Walter}]{Faherty2010}
Faherty, J.~K., Burgasser, A.~J., West, A.~A., Bochanski, J.~J., Cruz, K.~L.,
  Shara, M.~M., \& Walter, F.~M. 2010, The Astronomical Journal, 139, 176

\bibitem[{Geballe {et~al.}(2001)Geballe, Saumon, Leggett, Knapp, Marley, \&
  Lodders}]{Geballe2001}
Geballe, T.~R., Saumon, D., Leggett, S.~K., Knapp, G.~R., Marley, M.~S., \&
  Lodders, K. 2001, The Astrophysical Journal, 556, 373

\bibitem[{Goldman {et~al.}(2010)Goldman, Marsat, Henning, Clemens, \&
  Greiner}]{Goldman2010}
Goldman, B., Marsat, S., Henning, T., Clemens, C., \& Greiner, J. 2010, Monthly
  Notices of the Royal Astronomical Society, 405, 1140

\bibitem[{Gray {et~al.}(2006)Gray, Corbally, Garrison, McFadden, Bubar,
  McGahee, O'Donoghue, \& Knox}]{Gray2006}
Gray, R.~O., Corbally, C.~J., Garrison, R.~F., McFadden, M.~T., Bubar, E.~J.,
  McGahee, C.~E., O'Donoghue, A.~A., \& Knox, E.~R. 2006, The Astronomical
  Journal, 132, 161

\bibitem[{Groot {et~al.}(2009)Groot, Verbeek, Greimel, Irwin,
  Gonz\'{a}lez-Solares, G\"{a}nsicke, de~Groot, Drew, Augusteijn, Aungwerojwit,
  Barlow, Barros, {van Den Besselaar}, Casares, Corradi, Corral-Santana,
  Deacon, van Ham, Hu, Heber, Jonker, \& King}]{Groot2009}
Groot, P.~J., {et~al.} 2009, Monthly Notices of the Royal Astronomical Society,
  399, 323

\bibitem[{Hambly {et~al.}(2001)Hambly, MacGillivray, Read, Tritton, Thomson,
  Kelly, Morgan, Smith, Driver, Williamson, Parker, Hawkins, Williams, \&
  Lawrence}]{Hambly2001}
Hambly, N., {et~al.} 2001, Monthly Notices of the Royal Astronomical Society,
  326, 1279

\bibitem[{H\o~g {et~al.}(2000)H\o~g, Fabricius, Makarov, Urban, Corbin, Wycoff,
  Bastian, \& Schwekendiek}]{Hog2000}
H\o~g, E., Fabricius, C., Makarov, V.~V., Urban, S., Corbin, T., Wycoff, G.,
  Bastian, U., \& Schwekendiek, P. 2000, 30, 27

\bibitem[{Kaiser {et~al.}(2002)Kaiser, Aussel, Burke, Boesgaard, Chambers,
  Chun, Heasley, Hodapp, Hunt, Jedicke, Jewitt, Kudritzki, Luppino, Maberry,
  Magnier, Monet, Onaka, Pickles, Rhoads, Simon, Szalay, Szapudi, Tholen,
  Tonry, Waterson, \& Wick}]{Kaiser2002}
Kaiser, N., {et~al.} 2002, in Survey and Other Telescope Technologies and
  Discoveries. Proceedings of the SPIE, Volume 4836, ed. S.~{Tyson, J. Anthony;
  Wolff}, 154--164

\bibitem[{Kaiser {et~al.}(2010)Kaiser, Burgett, Chambers, Denneau, Heasley,
  Jedicke, Magnier, Morgan, Onaka, \& Tonry}]{Kaiser2010}
Kaiser, N., {et~al.} 2010, 77330E--77330E--14

\bibitem[{Kasper {et~al.}(2009)Kasper, Burrows, \& Brandner}]{Kasper2009}
Kasper, M., Burrows, A., \& Brandner, W. 2009, The Astrophysical Journal, 695,
  788

\bibitem[{King {et~al.}(2003)King, Villarreal, Soderblom, Gulliver, \&
  Adelman}]{King2003}
King, J.~R., Villarreal, A.~R., Soderblom, D.~R., Gulliver, A.~F., \& Adelman,
  S.~J. 2003, The Astronomical Journal, 125, 1980

\bibitem[{King {et~al.}(2010)King, McCaughrean, Homeier, Allard, Scholz, \&
  Lodieu}]{King2010}
King, R.~R., McCaughrean, M.~J., Homeier, D., Allard, F., Scholz, R.-D., \&
  Lodieu, N. 2010, Astronomy and Astrophysics, 510, A99

\bibitem[{Kirkpatrick(2005)}]{Kirkpatrick2005}
Kirkpatrick, J.~D. 2005, Annual Review of Astronomy and Astrophysics, 43, 195

\bibitem[{Koen {et~al.}(2010)Koen, Kilkenny, van Wyk, \& Marang}]{Koen2010}
Koen, C., Kilkenny, D., van Wyk, F., \& Marang, F. 2010, Monthly Notices of the
  Royal Astronomical Society, 403, 1949

\bibitem[{Leggett {et~al.}(2007)Leggett, Marley, Freedman, Saumon, Liu,
  Geballe, Golimowski, \& Stephens}]{Leggett2007}
Leggett, S.~K., Marley, M.~S., Freedman, R., Saumon, D., Liu, M.~C., Geballe,
  T.~R., Golimowski, D.~a., \& Stephens, D.~C. 2007, The Astrophysical Journal,
  667, 537

\bibitem[{Leggett {et~al.}(2010)Leggett, Saumon, Burningham, Cushing, Marley,
  \& Pinfield}]{Leggett2010}
Leggett, S.~K., Saumon, D., Burningham, B., Cushing, M.~C., Marley, M.~S., \&
  Pinfield, D.~J. 2010, The Astrophysical Journal, 720, 252

\bibitem[{Leggett {et~al.}(2008)Leggett, Saumon, Albert, Cushing, Liu, Luhman,
  Marley, Kirkpatrick, Roellig, \& Allers}]{Leggett2008}
Leggett, S.~K., {et~al.} 2008, The Astrophysical Journal, 682, 1256

\bibitem[{L\'{e}pine \& Shara(2005)}]{Lepine2005}
L\'{e}pine, S., \& Shara, M.~M. 2005, The Astronomical Journal, 129, 1483

\bibitem[{Liu {et~al.}(2008)Liu, Dupuy, \& Ireland}]{Liu2008}
Liu, M.~C., Dupuy, T.~J., \& Ireland, M.~J. 2008, The Astrophysical Journal,
  689, 436

\bibitem[{Liu {et~al.}(2010)Liu, Dupuy, \& Leggett}]{Liu2010}
Liu, M.~C., Dupuy, T.~J., \& Leggett, S.~K. 2010, The Astrophysical Journal,
  722, 311

\bibitem[{Liu {et~al.}(2007)Liu, Leggett, \& Chiu}]{Liu2007a}
Liu, M.~C., Leggett, S.~K., \& Chiu, K. 2007, The Astrophysical Journal, 660,
  1507

\bibitem[{Liu {et~al.}(2006)Liu, Leggett, Golimowski, Chiu, Fan, Geballe,
  Schneider, \& Brinkmann}]{Liu2006}
Liu, M.~C., Leggett, S.~K., Golimowski, D.~a., Chiu, K., Fan, X., Geballe,
  T.~R., Schneider, D.~P., \& Brinkmann, J. 2006, The Astrophysical Journal,
  647, 1393

\bibitem[{Liu {et~al.}(2011{\natexlab{a}})Liu, Deacon, Magnier, Aller, Bowler,
  Dupuy, Goldman, Burgett, Chambers, Hodapp, Kaiser, Kudritzki, Morgan, Price,
  Tonry, \& Wainscoat}]{Liu2011}
Liu, M.~C., {et~al.} 2011{\natexlab{a}}, ApJL submitted, astro-ph:1107.4608

\bibitem[{Liu {et~al.}(2011{\natexlab{b}})Liu, Delorme, Dupuy, Bowler, Albert,
  Artigau, Reyle, Forveille, \& Delfosse}]{Liu2011a}
---. 2011{\natexlab{b}}, ApJ accepted, astro-ph:1103.0014

\bibitem[{Lodieu {et~al.}(2007)Lodieu, Hambly, Jameson, \&
  Hodgkin}]{Lodieu2008}
Lodieu, N., Hambly, N.~C., Jameson, R.~F., \& Hodgkin, S.~T. 2007, Monthly
  Notices of the Royal Astronomical Society, 383, 1385

\bibitem[{Lucas {et~al.}(2008)Lucas, Hoare, Longmore, Schr\"{o}der, Davis,
  Adamson, Bandyopadhyay, de~Grijs, Smith, Gosling, Mitchison, G\'{a}sp\'{a}r,
  Coe, Tamura, Parker, Irwin, Hambly, Bryant, Collins, Cross, Evans,
  Gonzalez-Solares, Hodgkin, Lewis, Read, Riello, Sutorius, Lawrence, Drew,
  Dye, \& Thompson}]{LucasGPS}
Lucas, P.~W., {et~al.} 2008, Monthly Notices of the Royal Astronomical Society,
  391, 136

\bibitem[{Luhman {et~al.}(2011)Luhman, Burgasser, \& Bochanski}]{Luhman2011}
Luhman, K.~L., Burgasser, A.~J., \& Bochanski, J.~J. 2011

\bibitem[{Luhman {et~al.}(2007)Luhman, Patten, Marengo, Schuster, Hora, Ellis,
  Stauffer, Sonnett, Winston, Gutermuth, Megeath, Backman, Henry, Werner, \&
  Fazio}]{Luhman2007}
Luhman, K.~L., {et~al.} 2007, The Astrophysical Journal, 654, 570

\bibitem[{Magnier {et~al.}(2008)Magnier, Liu, Monet, \& Chambers}]{Magnier2008}
Magnier, E.~A., Liu, M., Monet, D.~G., \& Chambers, K.~C. 2008, Proceedings of
  the International Astronomical Union, 248, 553

\bibitem[{Mamajek \& Hillenbrand(2008)}]{Mamajek2008}
Mamajek, E.~E., \& Hillenbrand, L.~A. 2008, The Astrophysical Journal, 687,
  1264

\bibitem[{Mart\'{\i}n {et~al.}(1998)Mart\'{\i}n, Basri, Gallegos, Rebolo,
  Zapatero-Osorio, \& Bejar}]{Martin1998}
Mart\'{\i}n, E.~L., Basri, G., Gallegos, J.~E., Rebolo, R., Zapatero-Osorio,
  M.~R., \& Bejar, V. J.~S. 1998, The Astrophysical Journal, 499, L61

\bibitem[{McCaughrean {et~al.}(2004)McCaughrean, Close, Scholz, Lenzen, Biller,
  Brandner, Hartung, \& Lodieu}]{McCaughrean2004}
McCaughrean, M.~J., Close, L.~M., Scholz, R.-D., Lenzen, R., Biller, B.,
  Brandner, W., Hartung, M., \& Lodieu, N. 2004, Astronomy and Astrophysics,
  413, 1029

\bibitem[{Middelkoop(1982)}]{Middelkoop1982}
Middelkoop, F. 1982, Astronomy \& Astrophysics, 107, 31

\bibitem[{Monet {et~al.}(2003)Monet, Levine, Canzian, Ables, Bird, Dahn,
  Guetter, Harris, Henden, Leggett, Levison, Luginbuhl, Martini, Monet, Munn,
  Pier, Rhodes, Riepe, Sell, Stone, Vrba, Walker, Westerhout, Brucato, Reid,
  Schoening, Hartley, Read, \& Tritton}]{Monet2003}
Monet, D.~G., {et~al.} 2003, The Astronomical Journal, 125, 984

\bibitem[{Moraux {et~al.}(2003)Moraux, Bouvier, Stauffer, \&
  Cuillandre}]{Moraux2003}
Moraux, E., Bouvier, J., Stauffer, J.~R., \& Cuillandre, J.-C. 2003, Astronomy
  and Astrophysics, 400, 891

\bibitem[{Mugrauer {et~al.}(2006)Mugrauer, Seifahrt, Neuh\"{a}user, \&
  Mazeh}]{Mugrauer2006}
Mugrauer, M., Seifahrt, A., Neuh\"{a}user, R., \& Mazeh, T. 2006, Monthly
  Notices of the Royal Astronomical Society: Letters, 373, L31

\bibitem[{Murray {et~al.}(2011)Murray, Burningham, Jones, Pinfield, Lucas,
  Leggett, Tinney, Day-Jones, Weights, Lodieu, {P\'{e}rez Prieto}, Nickson,
  Zhang, Clarke, Jenkins, \& Tamura}]{Murray2011}
Murray, D.~N., {et~al.} 2011, Monthly Notices of the Royal Astronomical
  Society, 414, 575

\bibitem[{Nakajima {et~al.}(1995)Nakajima, Oppenheimer, Kulkarni, Golimowski,
  Matthews, \& Durrance}]{Nakajima1995}
Nakajima, T., Oppenheimer, B.~R., Kulkarni, S.~R., Golimowski, D.~A., Matthews,
  K., \& Durrance, S.~T. 1995, Nature, 378, 463

\bibitem[{Noyes {et~al.}(1984)Noyes, Hartmann, Baliunas, Duncan, \&
  Vaughan}]{Noyes1984}
Noyes, R.~W., Hartmann, L.~W., Baliunas, S.~L., Duncan, D.~K., \& Vaughan,
  A.~H. 1984, The Astrophysical Journal, 279, 763

\bibitem[{Perryman {et~al.}(1997)Perryman, Lindegren, Kovalevsky, Hoeg,
  Bastian, Bernacca, Cr\'{e}z\'{e}, Donati, Grenon, van Leeuwen, van~der Marel,
  Mignard, Murray, {Le Poole}, Schrijver, Turon, Arenou, Froeschl\'{e}, \&
  Petersen}]{Perryman1997}
Perryman, M. A.~C., {et~al.} 1997, Astronomy \& Astrophysics, 323, L49

\bibitem[{Perryman {et~al.}(1998)Perryman, Brown, Lebreton, Gomez, Turon,
  {Cayrel de Strobel}, Mermilliod, Robichon, Kovalevsky, \&
  Crifo}]{Perryman1998}
---. 1998, Astronomy \& Astrophysics, 331, 81

\bibitem[{Pinfield {et~al.}(2006)Pinfield, Jones, Lucas, Kendall, Folkes,
  Day-Jones, Chappelle, \& Steele}]{Pinfield2006}
Pinfield, D.~J., Jones, H. R.~a., Lucas, P.~W., Kendall, T.~R., Folkes, S.~L.,
  Day-Jones, a.~C., Chappelle, R.~J., \& Steele, I.~a. 2006, Monthly Notices of
  the Royal Astronomical Society, 368, 1281

\bibitem[{Rayner {et~al.}(2003)Rayner, Toomey, Onaka, Denault, Stahlberger,
  Vacca, Cushing, \& Wang}]{Rayner2003}
Rayner, J.~T., Toomey, D.~W., Onaka, P.~M., Denault, A.~J., Stahlberger, W.~E.,
  Vacca, W.~D., Cushing, M.~C., \& Wang, S. 2003, Publications of the
  Astronomical Society of the Pacific, 115, 362

\bibitem[{Rieke {et~al.}(2008)Rieke, Blaylock, Decin, Engelbracht, Ogle,
  Avrett, Carpenter, Cutri, Armus, Gordon, Gray, Hinz, Su, \&
  Willmer}]{Rieke2008}
Rieke, G.~H., {et~al.} 2008, The Astronomical Journal, 135, 2245

\bibitem[{Santos {et~al.}(2005)Santos, Israelian, Mayor, Bento, Almeida, Sousa,
  \& Ecuvillon}]{Santos2005}
Santos, N.~C., Israelian, G., Mayor, M., Bento, J.~P., Almeida, P.~C., Sousa,
  S.~G., \& Ecuvillon, A. 2005, Astronomy and Astrophysics, 437, 1127

\bibitem[{Saumon \& Marley(2008)}]{Saumon2008}
Saumon, D., \& Marley, M.~S. 2008, The Astrophysical Journal, 689, 1327

\bibitem[{Schmitt {et~al.}(1995)Schmitt, Fleming, \& Giampapa}]{Schmitt1995}
Schmitt, J. H. M.~M., Fleming, T.~A., \& Giampapa, M.~S. 1995, The
  Astrophysical Journal, 450, 392

\bibitem[{Scholz(2010)}]{Scholz2010B}
Scholz, R.-D. 2010, Astronomy and Astrophysics, 510, L8

\bibitem[{Scholz {et~al.}(2003)Scholz, McCaughrean, Lodieu, \&
  Kuhlbrodt}]{Scholz2003}
Scholz, R.-D., McCaughrean, M.~J., Lodieu, N., \& Kuhlbrodt, B. 2003, Astronomy
  and Astrophysics, 398, L29

\bibitem[{Skrutskie {et~al.}(2006)Skrutskie, Cutri, Stiening, Weinberg,
  Schneider, Carpenter, Beichman, Capps, Chester, Elias, Huchra, Liebert,
  Lonsdale, Monet, Price, Seitzer, Jarrett, Kirkpatrick, Gizis, Howard, Evans,
  Fowler, Fullmer, Hurt, Light, Kopan, Marsh, McCallon, Tam, {Van Dyk}, \&
  Wheelock}]{Skrutskie2006}
Skrutskie, M.~F., {et~al.} 2006, The Astronomical Journal, 131, 1163

\bibitem[{Slesnick {et~al.}(2006)Slesnick, Carpenter, Hillenbrand, \&
  Mamajek}]{Slesnick2006}
Slesnick, C.~L., Carpenter, J.~M., Hillenbrand, L.~A., \& Mamajek, E.~E. 2006,
  The Astronomical Journal, 132, 2665

\bibitem[{Soderblom(1985)}]{Soderblom1985}
Soderblom, D.~R. 1985, The Astronomical Journal, 90, 2103

\bibitem[{Soderblom {et~al.}(1991)Soderblom, Duncan, \&
  Johnson}]{Soderblom1991}
Soderblom, D.~R., Duncan, D.~K., \& Johnson, D. R.~H. 1991, The Astrophysical
  Journal, 375, 722

\bibitem[{Soderblom \& Mayor(1993)}]{Soderblom1993}
Soderblom, D.~R., \& Mayor, M. 1993, The Astronomical Journal, 105, 226

\bibitem[{Stubbs {et~al.}(2010)Stubbs, Doherty, Cramer, Narayan, Brown, Lykke,
  Woodward, \& Tonry}]{Stubbs2010}
Stubbs, C.~W., Doherty, P., Cramer, C., Narayan, G., Brown, Y.~J., Lykke,
  K.~R., Woodward, J.~T., \& Tonry, J.~L. 2010, The Astrophysical Journal
  Supplement Series, 191, 376

\bibitem[{Thalmann {et~al.}(2009)Thalmann, Carson, Janson, Goto, McElwain,
  Egner, Feldt, Hashimoto, Hayano, Henning, Hodapp, Kandori, Klahr, Kudo,
  Kusakabe, Mordasini, Morino, Suto, Suzuki, \& Tamura}]{Thalmann2010}
Thalmann, C., {et~al.} 2009, The Astrophysical Journal, 707, L123

\bibitem[{Tokunaga {et~al.}(2002)Tokunaga, Simons, \& Vacca}]{Tokunaga2002}
Tokunaga, A.~T., Simons, D.~A., \& Vacca, W.~D. 2002, Publications of the
  Astronomical Society of the Pacific, 114, 180

\bibitem[{Tonry(2011)}]{Tonry2011}
Tonry, J. 2011, in prep.

\bibitem[{Vacca {et~al.}(2003)Vacca, Cushing, \& Rayner}]{Vacca2003}
Vacca, W.~D., Cushing, M.~C., \& Rayner, J.~T. 2003, Publications of the
  Astronomical Society of the Pacific, 115, 389

\bibitem[{van Leeuwen(2007)}]{vanLeeuwen2007}
van Leeuwen, F. 2007, Astronomy and Astrophysics, 474, 653

\bibitem[{Vaughan {et~al.}(1978)Vaughan, Preston, \& Wilson}]{Vaughan1978}
Vaughan, A.~H., Preston, G.~W., \& Wilson, O.~C. 1978, Publications of the
  Astronomical Society of the Pacific, 90, 267

\bibitem[{Voges {et~al.}(2000)Voges, Aschenbach, Boller, Brauninger, Briel,
  Burkert, Dennerl, Englhauser, Gruber, Haberl, Hartner, Hasinger, Pfeffermann,
  Pietsch, Predehl, Schmitt, Trumper, \& Zimmermann}]{Voges2000}
Voges, W., {et~al.} 2000, IAU Circ., 7432, 3

\bibitem[{Wang {et~al.}(2010)Wang, Protopapas, Chen, Alcock, Burgett, Dombeck,
  Grav, Morgan, Price, \& Tonry}]{Wang2009}
Wang, J.-H., {et~al.} 2010, The Astronomical Journal, 139, 2003

\bibitem[{Wilson {et~al.}(2001)Wilson, Kirkpatrick, Gizis, Skrutskie, Monet, \&
  Houck}]{Wilson2001}
Wilson, J.~C., Kirkpatrick, J.~D., Gizis, J.~E., Skrutskie, M.~F., Monet,
  D.~G., \& Houck, J.~R. 2001, The Astronomical Journal, 122, 1989

\bibitem[{Wilson(1963)}]{Wilson1963}
Wilson, O.~C. 1963, The Astrophysical Journal, 138, 832

\bibitem[{Wright {et~al.}(2010)Wright, Eisenhardt, Mainzer, Ressler, Cutri,
  Jarrett, Kirkpatrick, Padgett, McMillan, Skrutskie, Stanford, Cohen, Walker,
  Mather, Leisawitz, Gautier, McLean, Benford, Lonsdale, Blain, Mendez, Irace,
  Duval, Liu, Royer, Heinrichsen, Howard, Shannon, Kendall, Walsh, Larsen,
  Cardon, Schick, Schwalm, Abid, Fabinsky, Naes, \& Tsai}]{Wright2010}
Wright, E.~L., {et~al.} 2010, The Astronomical Journal, 140, 1868

\bibitem[{Wright {et~al.}(2004)Wright, Marcy, Butler, \& Vogt}]{Wright2004}
Wright, J.~T., Marcy, G.~W., Butler, R.~P., \& Vogt, S.~S. 2004, The
  Astrophysical Journal Supplement Series, 152, 261

\end{thebibliography}

\end{document}